\documentclass[aps,prl,twocolumn,superscriptaddress]{revtex4-2}

\usepackage{amsmath,amssymb,amsthm}
\usepackage{microtype}
\usepackage{unicode}
\usepackage[hidelinks,colorlinks,allcolors=blue]{hyperref}
\usepackage[capitalize]{cleveref}
\usepackage{graphicx}
\graphicspath{{./images/}}
\usepackage{mathtools}
\usepackage{dsfont}
\usepackage{relsize}
\usepackage[caption=false]{subfig}
\usepackage{xspace}
\usepackage{tikz}
\usepackage{bm}

\newcommand{\tr}[1]{\ensuremath{\mathrm{tr}\left[#1\right]}} 
\newcommand{\oprod}[2]{\ensuremath{\left(#1 \mid #2 \right)}} 
\newcommand{\oket}[1]{\ensuremath{|#1)}} 
\newcommand{\wt}[1]{\widetilde{#1}} 
\newcommand{\diff}{\ensuremath{\mathrm{d}\xspace}}
\newcommand{\liou}{\ensuremath{\mathcal{L}\xspace}}
\DeclareMathOperator{\spn}{span}

\DeclareMathOperator{\cosech}{cosech}

\newcommand{\subfigimg}[3][,]{%
  \setbox1=\hbox{\includegraphics[#1]{#3}}
  \leavevmode\rlap{\usebox1}
  \rlap{\hspace*{-3pt}\raisebox{\dimexpr\ht1-0.5\baselineskip}{\textbf{#2}}}
  \phantom{\usebox1}
}

\begin{document}

\title{Probing hydrodynamic crossovers with dissipation-assisted operator evolution}

\author{N.~S.~Srivatsa}
\thanks{These authors contributed equally.}
\affiliation{Department of Physics, King's College London, United Kingdom}

\author{Oliver Lunt}
\thanks{These authors contributed equally.}
\affiliation{Department of Physics, King's College London, United Kingdom}

\author{Tibor Rakovszky}
\affiliation{Department of Physics, Stanford University, Stanford, California 94305, USA}

\author{Curt von Keyserlingk}
\affiliation{Department of Physics, King's College London, United Kingdom}

\date{\today}

\begin{abstract}
    
Using artificial dissipation to tame entanglement growth, we chart the emergence of diffusion in a generic interacting lattice model for varying U(1) charge densities. We follow the crossover from ballistic to diffusive transport above a scale set by the scattering length, finding the intuitive result that the diffusion constant scales as $D \propto 1/\rho$ at low densities $\rho$.
Our numerical approach generalizes the Dissipation-Assisted Operator Evolution (DAOE) algorithm: in the spirit of the BBGKY hierarchy, we effectively approximate non-local operators by their ensemble averages, rather than discarding them entirely. This greatly reduces the operator entanglement entropy, while still giving accurate predictions for diffusion constants across all density scales.
We further construct a minimal model for the transport crossover, yielding charge correlation functions which agree well with our numerical data.
Our results clarify the dominant contributions to hydrodynamic correlation functions of conserved densities, and serve as a guide for generalizations to low temperature transport.
\end{abstract}

\maketitle

\noindent\textit{Introduction.}---The complexity of a many-body system can give rise to emergent, universal behavior. A principal example is hydrodynamics, which is an effective theory for the transport of conserved quantities at mesoscopic scales~\cite{forsterHydrodynamicFluctuationsBroken2019,liuLecturesNonequilibriumEffective2018a}. Consider, for example, a clean system of interacting quantum particles. Between collisions, the particles travel ballistically and coherently. But collisions are expected to randomize the particle motion (at sufficiently high particle/energy density) so that diffusive hydrodynamics emerges at scales set by the interparticle separation.
Although this scenario includes fluids with momentum conservation and Navier-Stokes hydrodynamics, diffusion is also expected to generically emerge above a crossover lengthscale in high temperature non-integrable lattice systems, where momentum is not conserved and the only hydrodynamic modes with local conserved densities are energy and/or charge~\cite{mukerjeeStatisticalTheoryTransport2006}.
However, while this crossover picture is general and intuitive, it has not been verified in generic quantum systems either theoretically or numerically, except in special limits, e.g., in finely-tuned integrable models~\cite{gopalakrishnanKineticTheorySpin2019,denardisDiffusionGeneralizedHydrodynamics2019,karraschTransportPropertiesOnedimensional2014}. 

This Letter remedies the situation. We develop a new method for simulating hydrodynamics accurately at a range of fillings, and with it we are able  to chart in detail the crossover between rapidly diffusing behaviour at half filling and ballistic/coherent transport at low filling. We are moreover able to extract well-converged estimates for diffusion constants and correlation functions across the entire range of fillings, and verify the intuitive result that the diffusion constant $D$ is set by the inverse particle density at low filling, i.e. $D\propto 1/\rho$, where $\rho$ is the particle density. On the back of our data, we construct a minimal model for spacetime correlation functions which captures the crossover between ballistic and diffusive behaviors, and agrees well with our numerical simulations.

\begin{figure}[t]
    \centering
    \includegraphics[width=\columnwidth]{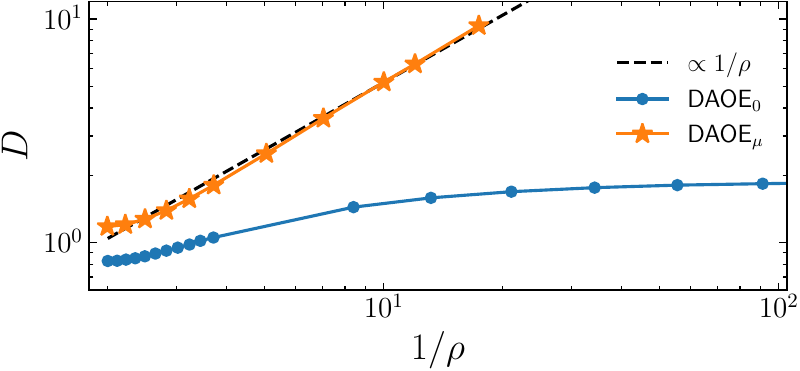}
    \caption{Comparison of the spin diffusion constant $D$ of the XX ladder against inverse density $\rho(\mu) = \frac{1}{2}(1 - \tanh{\frac{\mu}{2}})$, as calculated using the old DAOE$_0$ dissipator, and our new DAOE$_\mu$ dissipator. While the new dissipator gives the expected $D \propto 1/\rho$ scaling, the old dissipator does not correctly capture this behaviour, instead giving a $D(\rho)$ curve which saturates as $1/\rho \to \infty$. With $\gamma=0.5$, these data used $\ell_{*} = 3$ for DAOE$_{0}$ and $\ell_{*}=2$ for DAOE$_{\mu}$, hence the differing results at half filling $\rho=\frac{1}{2}$ (where DAOE$_{\mu}$ reduces to DAOE$_{0}$).}
    \label{fig:naive_daoe_diffusion}
\end{figure}

Any feasible classical simulation of a many body system needs to make approximations, and our work is no exception. A key technical result of this work is a theoretical understanding of how to approximate many-body dynamics at different fillings. Our numerical method is an extension of dissipation-assisted operator evolution (DAOE)~\cite{rakovszkyDissipationassistedOperatorEvolution2022,vonkeyserlingkOperatorBackflowClassical2022,lloydBallisticDiffusiveCrossover2023,kuoEnergyDiffusionWeakly2023}. As we will explain, the original method  approximates the true dynamics at lower computational cost by ignoring certain contributions to hydrodynamical correlators, namely those arising due to high-$n$-point functions. The intuition is that discarding high-$n$-point functions has little effect on hydrodynamics. Here we explain why this naive procedure fails at lower filling, leading to a premature onset of diffusive scaling (\cref{fig:naive_daoe_diffusion}). We also explain how a simple modification of the procedure---replacing high-$n$-point correlators by their disconnected components, rather than throwing them away entirely---fixes the issue while still reducing computational cost. In this sense, our generalized approximation scheme is similar in spirit to the approximation of the high-order correlations in the BBGKY hierarchy~\cite{bogolyubovLecturesQuantumStatistics1970,kardarStatisticalPhysicsParticles2007}.

\begin{figure}[t]
    \begin{minipage}[c][4.25cm][t]{0.52\columnwidth}
        \includegraphics[width=\columnwidth]{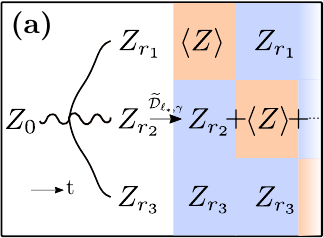}
    \end{minipage}
    \begin{minipage}{0.465\columnwidth}
        \includegraphics[width=\columnwidth]{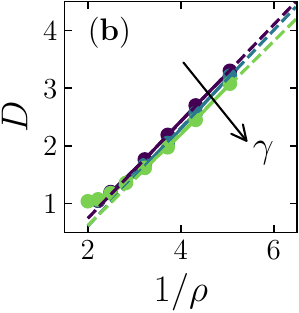}
    \end{minipage}
    \includegraphics[width=0.48\columnwidth]{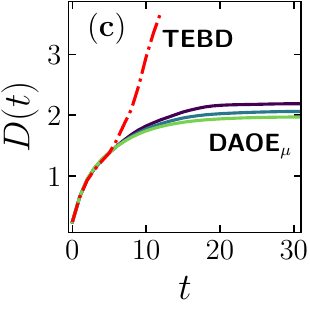}
    \hfill
    \includegraphics[width=0.507\columnwidth]{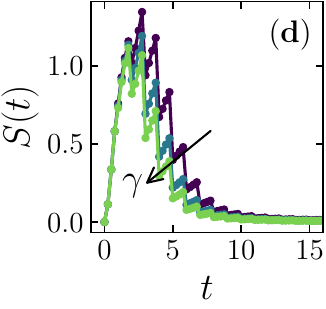}
    \caption{\textbf{(a)} Sketch of the artificial dissipation scheme DAOE$_{\mu}$. It replaces high-weight operators by lower-weight ones, with their contributions to hydrodynamic correlation functions approximated by a local ensemble average. \textbf{(b)} $D \propto 1/\rho$ scaling of the spin diffusion constant of the XX ladder, using the DAOE$_\mu$ dissipator $\wt{\mathcal{D}}_{\ell_{*},\gamma}$, with cutoff length $\ell_{*} = 2$ and dissipation strength $\gamma=0.1,0.15,0.2$ shown in different colors. Unlike in \cref{fig:naive_daoe_diffusion}, this new dissipator results in the expected scaling $D \sim 1/\rho$ when $1/\rho \gg 1$. \textbf{(c)} Comparison of the time-dependent diffusion constant $D(t)$ obtained using naive TEBD vs DAOE$_\mu$, with $\mu = 1$, $L = 201$, and $\chi_{\mathrm{max}} = 768$. At early times the methods agree, but after around $t \gtrsim 6$ truncation errors cause the TEBD estimate to become unstable, whereas DAOE$_\mu$ produces $D(t)$ curves that are well converged as $t \to \infty$. \textbf{(d)} The half-cut von Neumann operator entanglement entropy $S(t)$ of $\wt{\sigma}^{z}(t)$ time-evolved using DAOE$_\mu$. For finite dissipation strength $S(t)$ reaches an $\mathcal{O}(1)$ peak before decaying to zero, thus allowing for a faithful description with modest bond dimension. The staggering in the entropy is because we only apply the DAOE$_\mu$ dissipator every 5 timesteps.}
    \label{fig:D_exp_mu_scaling}
\end{figure}

\begin{figure}[t]
    \centering
    \subfigimg[width=\columnwidth]{(a)}{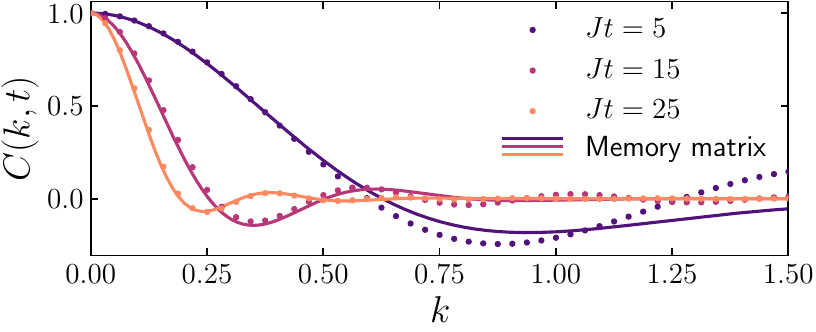}
    \subfigimg[width=\columnwidth]{(b)}{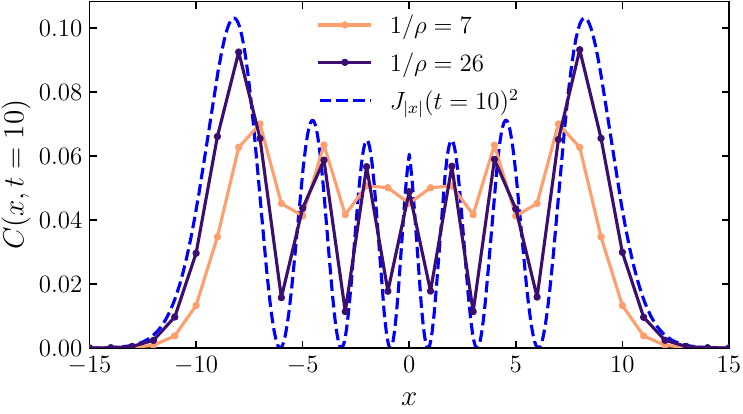}
    \caption{\textbf{(a)} Momentum space density-density connected correlation function $C(k,t) = \oprod{\wt{\sigma}^{z}_{k}(t)}{\wt{\sigma}^{z}_{k}}_{\mu}$, with a comparison between numerical data from DAOE$_\mu$ (dots), and the memory matrix prediction (lines), given in \cref{eq:mm_C1,eq:mm_C2}. Here $\mu = 1.4$, $L = 201$, $\ell_{*} = 2$, $\gamma = 0.2$, and $\chi_{\mathrm{max}} = 768$, and the diffusion constant is $D \approx 3.1$. We take the fit parameter for the memory matrix decay rate to be $\alpha = 2.5$. \textbf{(b)} Comparison of the real space density-density correlation function $C(x,t)$ at $t=10$ with the free-particle solution $J_{|x|}(t)^{2}$, where $J_{|x|}$ is a Bessel function. For a fixed $t$, as $1/\rho \to \infty$ the correlator becomes better described by the free solution, indicating the ballistic transport below the crossover scale $\sim 1/\rho$. Here $L = 121$, $\ell_{*} = 2$, $\gamma = 0.5$, and $\chi_{\mathrm{max}} = 512$.}
    \label{fig:kspace_correlations}
\end{figure}

\noindent\textit{Background.}---Various recent numerical methods have confirmed that diffusion emerges in lattice systems with a \emph{high} energy/particle density, where quasiparticles collide frequently~\cite{rakovszkyDissipationassistedOperatorEvolution2022,vonkeyserlingkOperatorBackflowClassical2022,lloydBallisticDiffusiveCrossover2023,kuoEnergyDiffusionWeakly2023,whiteQuantumDynamicsThermalizing2018,kleinkvorningTimeevolutionLocalInformation2022,artiacoEfficientLargeScaleManyBody2024,thomasComparingNumericalMethods2023,frias-perezConvertingLongrangeEntanglement2024,parkerUniversalOperatorGrowth2019}. Our task is to find a method that works well at a range of fillings. We do so by extending the original DAOE algorithm, designed to capture transport at infinite temperature/half filling, which we now explain.

For concreteness, consider a system of hardcore interacting bosons. This may be represented by a spin-$\frac{1}{2}$ chain of length $L$, with on-site particle density $n=(1+\sigma^z)/2$. All operators can be written as a superposition of the $4^{L}$ possible tensor products of the four Pauli matrices $\sigma^{\alpha=0,x,y,z}$. For a given `Pauli string' $\sigma^{\mathbf{\alpha}} \coloneqq \otimes_{i=1}^{L} \sigma_{i}^{\alpha_{i}}$, we define its \textit{weight} $\ell_{\alpha}$ to be the number of nonidentities in the product.

The connected charge density correlation function can be written as $C_{ij}(t) = \oprod{\sigma_{i}^{z}(t)}{\sigma_{j}^{z}}^{c}_{0}$ in the spin language. It plays a prominent role in hydrodynamics and encodes the diffusion constant. Here $\oprod{A}{B}_{0} \coloneqq \tr{A^{\dag}B} / \tr{\mathds{1}}$ is the infinite temperature thermal inner product.  By Trotterising the Heisenberg time evolution, we may express the correlator as a Feynman sum of paths $\mathcal{P}$ through the space of Pauli strings $C_{ij}(t)=\sum_{\mathcal{{P}}}W_\mathcal{P}$, where $W_\mathcal{P}$ is a Feynman path amplitude. The paths obey boundary conditions $\mathcal{P}(t=0)=\sigma^z_i,\mathcal{P}(t)=\sigma^z_j$, and so start and end at a pair of low weight operators. The idea underlying DAOE is that paths that involve high-weight Pauli strings may be discarded to good approximation. Heuristically, this is because there are exponentially more high weight operators than low weight operators, so under ergodic dynamics operators are more likely to continue to grow, rather than to shrink back down to a small operator like $\sigma^z_j$ (this argument can be made more precise~\cite{vonkeyserlingkOperatorBackflowClassical2022}). This motivates the introduction of \textit{artificial dissipation}, which suppresses high weight operators in order to reduce the operator entanglement, while minimizing the impact on hydrodynamics. The reduction in entanglement entropy reduces the bond dimension necessary to faithfully capture the time-evolved operator as a tensor network (in 1D), enabling simulations to longer times than would be possible without the dissipation. In more detail, we define the `DAOE$_0$' \textit{dissipation superoperator} $\mathcal{D}_{\ell_{*},\gamma}$ by its action on Pauli strings:
\begin{equation}
    \mathcal{D}_{\ell_{*}, \gamma}[\sigma^{\alpha}] \coloneqq \begin{cases}
        \sigma^{\alpha}, & \text{if } \ell_{\alpha} \leq \ell_{*};\\
        e^{-\gamma(\ell_{\alpha} - \ell_{*})} \sigma^{\alpha}, & \text{otherwise.}
    \end{cases}
\end{equation}
This exponentially suppresses all operators $\sigma^{\alpha}$ whose weight $\ell_{\alpha}$ is above some chosen cutoff $\ell_{*}$. Here $\gamma$ is a free parameter controlling the strength of the dissipation. If we treat the time-evolved operator as a matrix product state (MPS) in a doubled Hilbert space, then it turns out that the dissipator $\mathcal{D}_{\ell_{*},\gamma}$ can be implemented in a finite system using a matrix product operator (MPO) of bond dimension $\ell_{*} + 1$~\cite{rakovszkyDissipationassistedOperatorEvolution2022}. In practice we propose to apply the dissipator periodically with period $\Delta t$ as follows. Write $\oket{q(t)}$ for the vectorization of an operator $q(t)$, and $\mathcal{L} \coloneqq [H, \cdot \,]$ for the Liouvillian superoperator. Then the DAOE$_{0}$ modified time evolution in the time window $t \in [N \Delta t, (N+1)\Delta t)$ (for $N \in \mathbb{N}$) is defined by
\begin{equation}
    \oket{\tilde{q}(t)} \coloneqq e^{i \mathcal{L}(t - N \Delta t)} \left( \mathcal{D}_{\ell_{*}, \gamma} e^{i \mathcal{L} \Delta t}\right)^{N} \oket{q(0)}.
\end{equation}

\noindent\textit{Failure of the DAOE$_0$ dissipator at low filling.}---DAOE$_0$ performs well in the regime of infinite temperature and half-filling, yielding $\sim 1 \%$ accurate estimates of the diffusion constant for various models~\cite{rakovszkyDissipationassistedOperatorEvolution2022}. Now consider the limit of infinite temperature but finite particle density; for notational simplicity, we relabel $\beta \mu \to \mu$ throughout. In the limit $\mu \gg 1$, the equilibrium particle density $\rho = \frac{1}{2}(1 - \tanh{\frac{\mu}{2}}) \sim e^{-\mu}$ is very small. This implies that charge transport is ballistic up to times set by the interparticle spacing $1/\rho$, leading to the scaling $D \propto 1/\rho$. However, as we will now show, applying DAOE$_0$ dissipation at low-filling leads to a finite diffusion constant even as $1/\rho \to \infty$ (see \cref{fig:naive_daoe_diffusion}). 

The relevant operator inner product for computing equilibrium correlation functions is
\begin{equation}
    \oprod{A}{B}_{\mu} \coloneqq \tr{A^{\dag} \rho_{\mu}^{1/2} B \rho_{\mu}^{1/2}},
    \label{eq:finite_mu_prod}
\end{equation}
where $\rho_{\mu} = e^{-(\mu/2)\sigma^{z}_{\mathrm{tot}}} / \tr{e^{-(\mu/2)\sigma^{z}_{\mathrm{tot}}}}$ is the equilibrium density matrix. To understand the failure of DAOE$_{0}$ at low fillings, we expand $\rho_{\mu}$ in the parameter $\eta \coloneqq \langle \sigma^{z} \rangle_{\mu} = -\tanh(\mu/2)$, so that the $\mu \neq 0$ correlation function can be expressed as a sum of $\mu = 0$ correlators:
\begin{align}
    C(r,t) &= \oprod{\sigma^{z}_{r}}{\sigma^{z}_{0}(t)}_{\mu},\nonumber\\
    &= \sum_{k=0}^{\infty} \eta^{k} \Big(\sigma^{z}_{r} \, \left| \, [\mathcal{Z}^{(k)} \sigma^{z}_{0}](t)\right)_{\mu=0}, \label{eq:correlation_func_eta_expansion}
\end{align}
where $\mathcal{Z}^{(k)} \coloneqq \sum_{r_{1} < r_{2} < \cdots < r_{k}} \sigma^{z}_{r_{1}} \sigma^{z}_{r_{2}} \cdots \sigma^{z}_{r_{k}}$. Now we imagine including a round of DAOE$_{0}$ dissipation in the time evolution. $\mathcal{Z}^{(k)} \sigma^{z}_{0}$ is a $(k+1)$-body operator, and the entropic arguments that originally motivated DAOE$_{0}$ suggest that, when $k$ is large, there is only a small Feynman amplitude for $[\mathcal{Z}^{(k)} \sigma^{z}_{0}](t)$ to shrink down to a few-body operator~\cite{rakovszkyDissipationassistedOperatorEvolution2022,vonkeyserlingkOperatorBackflowClassical2022,nahumOperatorSpreadingRandom2018,khemaniOperatorSpreadingEmergence2018,vonkeyserlingkOperatorHydrodynamicsOTOCs2018,rakovszkyDiffusiveHydrodynamicsOutofTimeOrdered2018}. Therefore, performing DAOE$_{0}$ dissipation with a finite cutoff $\ell_{*}$ has the effect of approximately truncating this expansion at $k \sim \mathcal{O}(\ell_{*})$. We expect that the remaining $\mu=0$ correlators are generically diffusive at any finite order $k$, so their sum should remain diffusive, leading to a finite diffusion constant even as $\mu \to \infty$ (\cref{fig:naive_daoe_diffusion}). This finite order truncation is likely a particularly poor approximation at low fillings, since the higher order terms become less and less suppressed as $|\eta| \to 1$. This explains why the DAOE$_0$ dissipator erroneously gives diffusive behavior at low fillings, where the dynamics should instead be ballistic. In \cref{fig:daoe_default_diffusion} we numerically check this argument, finding close quantitative agreement at late times between $C(r,t)$ as computed using the full DAOE$_{0}$ evolution at \emph{finite} dissipation strength $\gamma$ and $\ell_{*}=3$, and the RHS of \cref{eq:correlation_func_eta_expansion} truncated at $k=3$. 

\noindent\textit{A new approximation scheme.}---The argument above suggests that  Pauli strings consisting solely of $\sigma^{z}$ or $\mathds{1}$ identity operators (or `$Z$-strings') become important at low filling, and that DAOE$_0$ fails because it throws away their important contributions to correlation functions. The same observation can be stated in terms of the thermal product in \cref{eq:finite_mu_prod}: unlike in the $\mu = 0$ case, long $Z$-strings can have non-vanishing inner products with short $Z$-strings. Indeed, these overlaps can be large at extreme fillings; for example, if $\alpha,\alpha^{\prime}$ are two $Z$-strings then $|(\sigma^\alpha|\sigma^{\alpha^{\prime}})_{\mu}|= |\eta|^{\ell_{\alpha+\alpha^{\prime}}}$, which tends to $1$ as $|\mu|\rightarrow\infty$. Therefore there are significant contributions to the correlator $\oprod{\sigma_{i}^{z}(t)}{\sigma_{j}^{z}}_{\mu}$ from terms in the Pauli-string expansion of $\sigma_{i}^{z}(t)=\sum_\alpha d_\alpha(t) \sigma^\alpha$ corresponding to high weight $Z$-strings. DAOE$_0$ incurs significant errors by suppressing these contributions. 

Fortunately, there is a simple fix which retains contributions from long $Z$-strings. The idea is to redefine our notion of the `weight' of an operator, taking into account the inner product \cref{eq:finite_mu_prod}. First we define an orthonormal on-site operator basis with respect to \cref{eq:finite_mu_prod} by Gram-Schmidt orthonormalizing   $\{\mathds{1}, \sigma^{z,x,y}\}$
\begin{align}
    \wt{\sigma}^{0} &= \mathds{1},\nonumber\\
    \wt{\sigma}^{z} &= c_{\mu}\left(\sigma^{z} - \langle \sigma^{z}\rangle_{\mu} \mathds{1}\right),\label{eq:tilde_pauli_def}\\
    \wt{\sigma}^{x,y} &= \sqrt{c_{\mu}} \sigma^{x,y}.\nonumber
\end{align}
Here $c_{\mu} = \cosh(\mu/2)$, and $\langle \sigma^{z} \rangle_{\mu} = -\tanh(\mu/2)$ is the equilibrium expectation value of $\sigma^{z}$. We can consider operator strings in this new basis, $\wt{\sigma}^{\mathbf{\alpha}} = \otimes_{i=1}^{L} \wt{\sigma}^{\alpha_{i}}$, and we redefine the `operator weight' to be the number of terms which are not $\wt{\sigma}^{0}$ appearing in this product. We remark that the nature of this dissipation is reminiscent of the BBGKY approximation~\cite{bogolyubovLecturesQuantumStatistics1970,kardarStatisticalPhysicsParticles2007}, in the sense that `discarding' the operator $\wt{\sigma}^{z} \mapsto 0$ amounts to replacing $\sigma^{z}$ with its ensemble expectation value $\langle \sigma^{z} \rangle_{\mu}$. Thus, while the new dissipator still acts nontrivially on long $Z$-strings, it is able to approximately capture their contributions to correlation functions by pushing their weight on to shorter $Z$-strings (see \cref{sec:operator_weight}), whereas the old DAOE$_{0}$ dissipator would discard them entirely.  

Dissipating operators which are high weight with respect to this new definition can be performed by a straightforward modification of the DAOE$_0$ dissipator. We simply perform a change of basis before and after applying the DAOE$_{0}$ dissipator, so that the action of this `DAOE$_{\mu}$' dissipator on a Pauli string $\sigma^{\alpha}$ is
\begin{equation}
    \wt{\mathcal{D}}_{\ell_{*}, \gamma}[\sigma^{\alpha}] \coloneqq M_{\mu} \mathcal{D}_{\ell_{*}, \gamma}[M_{\mu}^{-1} \sigma^{\alpha}],
\end{equation}
where $M_{\mu}$ is the (non-unitary) matrix changing from the basis $\{\wt{\sigma}^{0}, \wt{\sigma}^{z,x,y}\}$ to the basis $\{\sigma^{0}, \sigma^{z,x,y}\}$. Since this change of basis is on-site, $\wt{\mathcal{D}}_{\ell_{*},\gamma}$ has the same bond dimension $\ell_{*}+1$ as $\mathcal{D}_{\ell_{*}, \gamma}$. 

We benchmarked this new DAOE$_{\mu}$ dissipator by studying spin transport in the XX ladder
\begin{align}
    H = &\frac{J}{4}\sum_{r=-(L-1)/2}^{(L-1)/2} \sum_{a=1,2} \left(\sigma^{x}_{r,a} \sigma^{x}_{r+1,a} + \sigma^{y}_{r,a} \sigma^{y}_{r+1,a}\right) \nonumber\\
    + &\frac{J}{4}\sum_{r=-(L-1)/2}^{(L-1)/2} \left(\sigma^{x}_{r,1}\sigma^{x}_{r,2} + \sigma^{y}_{r,1}\sigma^{y}_{r,2}\right),
\end{align}
where $a=1,2$ labels the legs of the ladder, and we set $J = 1$ and take $L$ odd. This model conserves both energy and total $\sigma^{z}$, and is non-integrable due to the ladder coupling, leading to diffusive spin transport~\cite{steinigewegScalingDiffusionConstants2014,karraschSpinThermalConductivity2015,klossTimedependentVariationalPrinciple2018,znidaricCoexistenceDiffusiveBallistic2013}. We compute the spin diffusion constant $D = \lim_{t \to \infty} D(t)$ using $D(t) = \partial_{t} \sum_{r}\frac{r^{2}}{2} C(r,t)$, where $C(r,t) = \oprod{\sigma^{z}_{r}(t)}{\sigma^{z}_{0}}^{c}_{\mu}$ is the connected correlation function for $\sigma^{z}_{r} = (\sigma^{z}_{r,1} + \sigma^{z}_{r,2})/2$, normalized to $\sum_{r} C(r,t) = 1$. At low particle densities $\rho = \frac{1}{2}(1 - \tanh{\frac{\mu}{2}}) \ll 1$, one expects that $D \propto 1/\rho$, which signals the onset of ballistic transport on lengthscales smaller than the typical interparticle spacing $1/\rho$~\cite{sachdevQuantumPhaseTransitions2011}. Whereas this was not present with the old DAOE$_{0}$ dissipator (c.f.~\cref{fig:naive_daoe_diffusion}), we can see from \cref{fig:D_exp_mu_scaling} that the DAOE$_\mu$ dissipator recovers the expected $D \propto 1/\rho$ scaling.  For a more detailed check, in \cref{fig:kspace_correlations}b we show the connected correlator $C(r,t)$ for different densities $\rho$. This shows increasingly good agreement with the expected single-particle solution $J_{|r|}(t)^{2}$ as $1/\rho \to \infty$, again reflecting ballistic transport on scales below $1/\rho$.

We are able to get converged estimates of diffusion constants $D$ down to $\gamma = 0.1$ and up to $1/\rho \approx 5$; this is increasingly challenging as $1/\rho$ increases because $D(t)$ saturates on an $\mathcal{O}(1/\rho)$ timescale. The need for these long timescales can cause standard methods like TEBD to break down before giving converged estimates for $D$ (see inset to \cref{fig:D_exp_mu_scaling}). The advantage of using DAOE$_{\mu}$ is that it allows for a numerically controlled calculation down to some finite $\gamma$, since the artificial dissipation limits the operator entanglement growth (see inset to \cref{fig:kspace_correlations}). Assuming $D$ depends smoothly on $\gamma$, one can then extrapolate to $\gamma \to 0$ in order to estimate the true diffusion constant. In \cref{fig:dvsgam} we show this extrapolation for $\ell_{*} = 2, 3$, with both cutoffs giving $\gamma \to 0$ extrapolations that agree to within $\lesssim 10 \%$.

\noindent \textit{Modeling the crossover.}---Inspired by the success of our numerical method in capturing the ballistic-to-diffusive crossover of charge correlations above an $\mathcal{O}(1/\rho)$ scale, we  construct a simplified analytical model which \emph{qualitatively} captures the same behaviour. We use the memory matrix formalism~\cite{forsterHydrodynamicFluctuationsBroken2019}, which requires that we identify a space of `slow' operators; operators not in the slow space are called `fast'. Since we are interested in density-density correlation functions $\oprod{\wt{\sigma}^{z}_{r}(t)}{\wt{\sigma}^{z}_{0}}$, we include $\wt{\sigma}^{z}$ in the slow space. We also choose to include in the slow space all weight $\ell=1,2$ operators that could possibly appear in $\wt{\sigma}^{z}_{r}(t)$ while respecting the symmetries of the XX ladder, and consign all $(\ell \geq 3)$ operators to the fast space.

In order to close the equations of motion for the slow operators, we need to postulate a form for the correlation functions of fast operators, which are expected to decay rapidly. We will assume this in a very strong sense, making the uncontrolled approximation
\begin{equation}
    (f_{k,a} | e^{-i Q \liou Q t} | f_{k,b} )_{\mu} \approx \lambda \delta_{a b} \delta(t),
    \label{eq:mm_approx}
\end{equation}
where $f_{k,a}$, $f_{k,b}$ denote momentum space fast operators with `flavor indices' $a,b$ whose explicit form is unimportant here~\cite{SM}, $Q$ is the projector on to the fast space, and $\lambda$ is a constant. With this approximation made, we perform a standard memory matrix computation, finding the appearance of a dissipation rate  $\Gamma \equiv \lambda c_{\mu}^{-2}$ which relates to the diffusion constant as $D = (2\Gamma)^{-1}$~\cite{SM}. Thus the main free parameter in our model, $\lambda$, is fixed by the value of the diffusion constant. Since $\Gamma \sim \lambda e^{-|\mu|}$, if $\lambda$ has only weak (polynomial) dependence on $\mu$, then $D$ grows exponentially with $\mu$, as expected.

Within this approximation for the fast space dynamics, we find that the desired density-density correlation function $C(k,t) \coloneqq \oprod{\wt{\sigma}^{z}_{k}(t)}{\wt{\sigma}^{z}_{k}}_{\mu}$ decomposes into two parts, $C(k,t) = C_{1}(k,t) + C_{2}(k,t)$, each coming from different poles in the Green's function~\cite{SM}. $C_{1}$ is given by
\begin{equation}
    C_{1}(k,t) = \int_{-\pi}^{\pi} \dfrac{\diff q}{2\pi} \dfrac{2 \beta \eta^{2} \sin(q) e^{-\Gamma t + 2 i \beta t \cos{q}}}{-\Gamma \eta^{2} + 2 \beta \sin{q} - (1-\eta^{2})2i \beta \cos{q}}.
    \label{eq:mm_C1}
\end{equation}
where $\eta \coloneqq -\tanh(\mu/2)$ and $\beta \coloneqq -\eta \sin(k/2)$. This is the term which gives rise to ballistic physics. Indeed, in the $|\mu| \to \infty$ limit, $C_{1}$ reduces to $C_{1}(k,t) = J_{0}(2 t \sin(k/2))$, where $J_{0}(z)$ is a Bessel function, which is precisely the free-particle coherent behavior expected in the extremely dilute regime. 

$C_{2}$ results in the diffusive behavior, and is given by
\begin{equation}
    C_{2}(k,t) = \dfrac{\sum_{a=\pm} a \, e^{z_{a} t}(\Gamma + (1-\eta^{2})z_{a}) \, \mathcal{I}\left(\left|\frac{z_{a}}{2 \beta \eta^{-2}}\right|<1\right)}{\sqrt{\Gamma^{2} - 4\beta^{2}(2 \eta^{-2}-1)}},
    \label{eq:mm_C2}
\end{equation}
where $z_{\pm} = (1/(2-\eta^{2})) (-\Gamma \pm \sqrt{\Gamma^{2} - 4 \beta^{2} (2 \eta^{-2}-1)})$ and $\mathcal{I}$ is an indicator function. As $k \to 0$, $\beta \approx -\eta k/2$ and we recover the familiar diffusive scaling $C_{2}(k,t) \approx \exp(-D k^{2} t)$ with $D = (2\Gamma)^{-1}$. This scaling holds in the momentum regime $|k| \ll |\eta \Gamma| \sim e^{-|\mu|} \sim \rho$, i.e., on length scales larger than the typical interparticle spacing $1/\rho$.

Thus far, the only free parameter in our model, $\lambda$, was fixed by the diffusion constant via $D = (2\Gamma)^{-1} = (2\lambda c_{\mu}^{-2})^{-1}$. This effectively captures the late time, $k \to 0$ behavior. However, it is plausible that the decay rate of fast operators $f_{k,a}$ depends on $k$, and it is important to account for this to better quantitatively capture the finite $k$ correlation function. Aiming to strike a balance between accuracy and predictive power, we slightly refine the approximation in \cref{eq:mm_approx} and introduce some weak $k$-dependence into $\lambda$ by Taylor expanding
\begin{equation}
    \lambda(k) \approx \lambda_{0} + \alpha \sin^{2}{\frac{k}{2}},
    \label{eq:lambda_taylor_expansion}
\end{equation}
where now $\lambda_{0}$ is fixed by the diffusion constant via $D = (2\lambda_{0} c_{\mu}^{-2})^{-1}$, and $\alpha$ is a free parameter. The expansion parameter is $\sin(k/2)$ because we are on the lattice. \cref{fig:kspace_correlations}a shows a comparison between the memory matrix prediction \cref{eq:mm_C1,eq:mm_C2}, and numerical data from DAOE$_\mu$, with good quantitative agreement at small to moderate $k$. Note that the single fit parameter $\alpha$, here set to $\alpha = 2.5$, is time-independent, so it is nontrivial that we can match the numerical data over different timescales. We could likely improve the agreement at larger $k$ by including more Taylor terms in \cref{eq:lambda_taylor_expansion}, but even this low-order expansion gives reasonable results.

\noindent \textit{Conclusion.}---We introduced a numerical algorithm which enables the simulation of charge transport over the range of densities and timescales needed to resolve a crossover between ballistic and diffusive behaviour. Our method is a generalization of DAOE$_{0}$, and somewhat related to the BBGKY hierarchy.  We also derived a minimal analytical model which correctly captures the $\mathcal{O}(1/\rho)$ scaling of the crossover lengthscale, and agrees well with our numerics at small momenta.

Our generalization of DAOE$_{0}$ was prompted by considering which operator histories give dominant contributions to hydrodynamic correlation functions, and in particular how this changes with the chemical potential, keeping the temperature infinite. However, for generic systems, transport at $T=\infty$ is diffusive. It would therefore be exciting to generalize these ideas to lower temperatures, allowing for the study of more exotic transport.

Another question relates to the classical computational complexity of transport simulations using DAOE$_{\mu}$. For $\mu=0$, Ref.~\cite{vonkeyserlingkOperatorBackflowClassical2022} argued that DAOE$_{0}$ requires only $\exp[\mathcal{O}(\log^{2}(1/\epsilon))]$ memory to estimate a diffusion constant to precision $\epsilon$, a superpolynomial improvement over the $\exp[\mathcal{O}(\mathrm{poly}(1/\epsilon))]$ scaling for brute-force time evolution. Does using DAOE$_{\mu}$ entail the same saving for $\mu \neq 0$ transport?

\noindent \textit{Acknowledgements.}---CvK is supported by a UKRI Future Leaders Fellowship MR/T040947/1. T.R. is supported in part by the Stanford Q-Farm Bloch Postdoctoral Fellowship in Quantum Science and Engineering.

\bibliography{bibliography}

\clearpage
\widetext
\begin{center}
\textbf{\large Supplemental Material for `Probing hydrodynamic crossovers with dissipation-assisted operator evolution'}
\end{center}
\makeatletter
\renewcommand{\c@secnumdepth}{0}
\makeatother

\setcounter{equation}{0}
\setcounter{figure}{0}
\setcounter{table}{0}
\setcounter{page}{1}
\makeatletter
\renewcommand{\theequation}{S\arabic{equation}}
\renewcommand{\thefigure}{S\arabic{figure}}
\renewcommand{\thesection}{S\arabic{section}}

\section{Details of the memory matrix calculation}
In this section we give more details of the memory matrix calculation we used to model the crossover between ballistic and diffusive transport as a function of the chemical potential $\mu$. The memory matrix formalism requires that we identify a space of slow variables, which we denote by $O_{\mathrm{slow}} = \mathrm{span}\left\{a_{i}\right\}$. Let $P$ denote the projector on to the slow space, and $Q = \mathds{1} - P$ its complement. Assume without loss of generality that $\oprod{a_{i}}{a_{j}} = \delta_{ij}$. Defining $C_{ij}(t) = \oprod{a_{i}(t)}{a_{j}}$ and the Laplace transform $C_{ij}(z) = \int_{0}^{\infty} e^{i z t} C_{ij}(t) \diff t$, we find~\cite{forsterHydrodynamicFluctuationsBroken2019}
\begin{equation}
    C_{ij}(z) = \left[\dfrac{1}{z\mathds{1} - \mathbf{\Omega} + i \mathbf{\Sigma}}\right]_{ij},
\end{equation}
where
\begin{align}
    \Omega_{ij} &= (a_{i} | \mathcal{L} | a_{j}),\\
    \Sigma_{ij} &= (a_{i} | \mathcal{L}Q \dfrac{i}{z - Q \mathcal{L} Q} Q \mathcal{L} | a_{j}),
\end{align}
where $\mathcal{L} = [H, \cdot]$ is the Liouvillian. So far everything is exact. To make progress, it is past this point that approximations are made. A typical approximation is to assume that $\mathbf{\Sigma}(z)$ is regular in $z$, which is akin to assuming that correlations in the fast space die off quickly in time. We will make an approximation of this form below.

Our goal is to model density-density connected correlation functions $\oprod{\sigma_{r}^{z}(t)}{\sigma_{0}^{z}}_{\mu}^{c} \equiv \oprod{\sigma_{r}^{z}(t) - \langle \sigma^{z} \rangle_{\mu}}{\sigma_{0}^{z} - \langle \sigma^{z} \rangle_{\mu}}_{\mu} = c_{\mu}^{-2} \oprod{\wt{\sigma}_{r}^{z}(t)}{\wt{\sigma}_{0}^{z}}_{\mu}$. Thus we will include $\wt{\sigma}_{r}^{z}$ as one of our slow variables. But to obtain useful results, we will ned to include further variables too. $\wt{\sigma}_{r}^{z}$ is a 1-body observable. The next simplest operators to include are 2-body operators. Using the $X_{\mathrm{glob}} = \prod_{j,a} \sigma^{x}_{j,a}$ global symmetry of the Hamiltonian, and the fact that $\wt{\sigma}^{z}_{r} = (\wt{\sigma}^{z}_{r,1} + \wt{\sigma}^{z}_{r,2})/2$ is $1 \leftrightarrow 2$ swap symmetric, it is straightforward to verify that all the one- and two-body operators that could possibly appear in $\wt{\sigma}^{z}_{r}(t)$ are of the form
\begin{equation}
    a_{r,\Delta,\nu} \coloneqq \dfrac{1}{2\sqrt{2}}\left[s^{+}_{r,\nu} s^{-}_{r+\Delta,\nu} - s^{-}_{r,\nu}s^{+}_{r+\Delta,\nu} - 4 c_{\mu} \eta \delta_{\Delta,0} \mathds{1} \right],
\end{equation}
where $c_{\mu} = \cosh(\mu/2)$, $\eta = -\tanh(\mu/2)$, $\nu$ can take the values $\nu = \pm 1$, and $s^{\pm}_{r,\nu}$ is a linear combination of the $\wt{\sigma}$ operators [c.f.~\cref{eq:tilde_pauli_def}] given by
\begin{equation}
    s^{\pm}_{r,\nu} \coloneqq \wt{\sigma}^{\pm}_{r,1} + \nu \, \wt{\sigma}^{\pm}_{r,2}.
\end{equation}
The operator $a_{r,\Delta,\nu}$ includes an insertion of $s^{+}$ at position $r$, and an insertion of $s^{-}$ a distance $\Delta$ away. In the case $\Delta = 0$, we have $a_{r,\Delta=0,\nu} = \sqrt{2} \wt{\sigma}^{z}_{r}$ regardless of $\nu$, so we only keep $a_{r,\Delta=0,+}$ in our basis and discard $a_{r,\Delta=0,-}$ to avoid double counting the operator.
Note that $s^{\pm}_{r,\nu} \to \nu s^{\pm}_{r,\nu}$ under a $1 \leftrightarrow 2$ swap operation. Note too that $a_{r,\Delta,\nu}$ are all antisymmetric under $X_{\mathrm{glob}}$ and unit normalized w.r.t.~the thermal inner product \cref{eq:finite_mu_prod}.
Our slow space is then 
\begin{equation}
    O_{\mathrm{slow}} = \spn\left[\{a_{r,0,+}\}_{r} \cup \{a_{r,\Delta,\nu}\}_{r,\Delta>0,\nu}\right].
\end{equation}
Next we will give the equations of motion of the slow operators, for which purpose it is easier to work in momentum space, with $a_{k,\Delta,\nu} \coloneqq \sum_{r} e^{-ik(r + \Delta/2)} a_{r,\Delta,\nu}$. Then the equations of motion $\partial_{t} a_{k,\Delta,\nu} = i \liou(a_{k,\Delta,\nu})$ are determined by
\begin{align}
    \liou(a_{k,0,+}) &= -v_{k}\left(a_{k,1,+} + a_{k,1,-}\right),\\
    \liou(a_{k,1,\nu}) &= v_{k}\left(a_{k,0,+} + \eta a_{k,2,\nu}\right) + c_{\mu}^{-1} f_{k,1,\nu},\\
    \liou(a_{k,\Delta\geq 2,\nu}) &= v_{k} \eta\left(a_{k,\Delta+1,\nu} - a_{k,\Delta-1,\nu}\right) + c_{\mu}^{-1} f_{k,\Delta,\nu},
\end{align}
where $v_{k} \coloneqq i \sin(k/2)$, and $f_{k,\Delta,\nu}$ are sums of fast operators (in the image of $Q$) with density independent norms. Note however that these operators are accompanied by factors of $1/\cosh(\mu/2)$ which is small at extreme fillings (i.e.~when $|\mu|$ is large). Thus our slow operators leak out into the fast space, but the rate at which they do so is suppressed at extreme fillings. We otherwise do not need the explicit forms of the $f$ operators in what proceeds.

\subsection{Fast space approximation}
$\mathbf{\Sigma}$ and $\mathbf{\Omega}$ are large matrices acting on the slow space labelled by the compound index $i = (\Delta,\nu)$. Let us adopt the convention that the 0th element of this basis is $a_{k,0,+}$. In that case note that, because $Q \mathcal{L} a_{k,0} = 0$, we have $\Sigma_{j0} = \Sigma_{0,j}$ for any value of the index $j$. The non-vanishing elements of $\mathbf{\Sigma}$ will look like
\begin{align}
    \Sigma_{(\Delta,\nu);(\Delta^{\prime},\nu^{\prime})} &= c_{\mu}^{-2} (f_{k,\Delta,\nu} | \dfrac{i}{z - Q \mathcal{L} Q} | f_{k,\Delta^{\prime},\nu^{\prime}}),\\
    &= c_{\mu}^{-2} \int_{0}^{\infty} e^{i z t} (f_{k,\Delta,\nu} | e^{-i Q \mathcal{L} Q t} | f_{k,\Delta^{\prime},\nu^{\prime}}) \diff t.
\end{align}
We expect, although we cannot control this approximation, that the above correlation function decays quickly in time regardless of filling. This will ensure that $\mathbf{\Sigma}$ is regular in $z$, and also that the primary $\mu$ dependence of $\mathbf{\Sigma}$ comes from the $1/c_{\mu}^{2}$ factor outside of the integrand. Locality of the dynamics moreover suggests that $\Sigma_{(\Delta,\nu),(\Delta^{\prime},\nu^{\prime})}$ decays in $|\Delta - \Delta^{\prime}|$. Motivated by these considerations, we will make the aggressive approximation that
\begin{equation}
    (f_{k,\Delta,\nu} | e^{-i Q \mathcal{L} Q t} | f_{k,\Delta^{\prime},\nu^{\prime}}) \approx \lambda \delta_{\Delta \Delta^{\prime}} \delta_{\nu \nu^{\prime}} \delta(t),
\end{equation}
where $\lambda$ is a constant that doesn't strongly depend on the density. As a result, we approximate
\begin{equation}
    \Sigma_{(\Delta,\nu);(\Delta^{\prime},\nu^{\prime})} \approx \Gamma \delta_{\Delta > 0}\delta_{\Delta \Delta^{\prime}} \delta_{\nu \nu^{\prime}}.
\end{equation}
where we defined $\Gamma = \lambda c_{\mu}^{-2}$. We will soon see that, within this approximation, the diffusion constant is given by $D = (2\Gamma)^{-1}$.

\subsection{Green's function}
To obtain the Green's function we need to calculate $\oprod{a_{k,0}(z)}{a_{k,0}} = i [z \mathds{1} - \mathbf{\Omega} + i \mathbf{\Sigma}]^{-1}_{00}$, having now made the approximation for our fast space dynamics. In the fixed $k$-sector operator basis ordered $\{a_{k,0}, a_{k,1,+}, a_{k,2,+}, \dots, a_{k,1,-}, a_{k,2,-}, \dots\}$, the structure of the $\Sigma$ and $\Omega$ matrices are
\begin{equation}
    \Sigma = \left(\begin{array}{c|ccc|cccc}
        0 & & & & & & \\ \hline
        & \Gamma & & & & & & \\
        & & \ddots & & & & & \\
        & & & & & & & \\ \hline
        & & & & \Gamma & & & \\
        & & & & & \ddots & & \\
        & & & & & & & 
    \end{array}\right),
    \qquad
    \Omega = \left(\begin{array}{c|ccc|cccc}
        0 & v_{k} & 0 & \cdots & v_{k} & 0 & \cdots \\ \hline
        -v_{k} & & & & & & & \\
        0 & & \Omega_{+} & & & & & \\
        \vdots & & & & & & & \\ \hline
        -v_{k} & & & & & & & \\
        0 & & & & & \Omega_{-} & & \\
        \vdots & & & & & & & 
    \end{array}\right),
\end{equation}
where the blocks $\Omega_{\pm}$ are given by
\begin{equation}
    \Omega_{\pm} = \eta v_{k} \begin{pmatrix}
        0 & -1 & 0 & \cdots \\
        1 & 0 & -1 & \\
        0 & 1 & & \ddots \\
        \vdots & & \ddots
    \end{pmatrix}.
\end{equation}
Let $A = \mathbf{\Omega} - i \mathbf{\Sigma}$, and define $P_{0} = |0\rangle \langle 0|$, $Q_{0} = \mathds{1} - P_{0}$ as the projectors on to $a_{k,0} \equiv |0\rangle$ and its complement in the slow space. Using a standard manipulation based on Dyson's identity~\cite{forsterHydrodynamicFluctuationsBroken2019}, we have
\begin{equation}
    \left[\dfrac{1}{z \mathds{1} - A}\right]_{00} = \dfrac{1}{z - \langle 0 | A Q_{0} \frac{1}{z \mathds{1} - Q_{0} A Q_{0}} Q_{0} A | 0 \rangle},
\end{equation}
where we used $\langle 0 | A | 0 \rangle = 0$. From the block structure, we can see that $Q_{0} A |0\rangle = -v_{k}\left(a_{k,1,+} + a_{k,1,-}\right)$. Also, after applying the projectors, $Q_{0}AQ_{0}$ is now block-diagonal, with the only nonzero entries given in $\Omega_{\pm}$, and therefore there is no longer any coupling between the $+$ and $-$ sectors. As a result, we have
\begin{equation}
    \left(a_{k,1,\pm}\left|\dfrac{1}{z \mathds{1} - Q_{0}AQ_{0}} \right| a_{k,1,\mp}\right) = 0,
\end{equation}
and because $\Omega_{+} = \Omega_{-}$ we have
\begin{equation}
    \left(a_{k,1,+}\left|\dfrac{1}{z \mathds{1} - Q_{0}AQ_{0}} \right| a_{k,1,+}\right) = \left(a_{k,1,-}\left|\dfrac{1}{z \mathds{1} - Q_{0}AQ_{0}} \right| a_{k,1,-}\right).
\end{equation}
We can hence invert the $\pm$ blocks separately with the same result, so that the Green's function can be expressed as
\begin{equation}
    \oprod{a_{k,0}(z)}{a_{k,0}} = \dfrac{i}{z + 2 v_{k}^{2} (M^{-1})_{11}},
\end{equation}
where the matrix $M$ is given by
\begin{equation}
    M = (z + i \Gamma) \mathds{1} - \eta v_{k} \sum_{n\geq 1} \left(|n+1\rangle\langle n| - \mathrm{h.c.}\right).
\end{equation}
Fortunately it is relatively easy to find the inverse. Note that we can transform this into a hopping problem via a simple onsite transformation of the form $M^{\prime} = i^{-\hat{x}} M i^{\hat{x}}$, so that
\begin{equation}
    M^{\prime} = (z + i \Gamma) \mathds{1} + \beta \sum_{n\geq 1} \left(|n+1\rangle\langle n| + \mathrm{h.c.}\right),
\end{equation}
where $\beta = i \eta v_{k}$ (note $\beta$ is real). We have $\left[(M^{\prime})^{-1}\right]_{11} = \left[M^{-1}\right]_{11}$.

This is currently a problem on the semi-infinite chain with sites $1,2,\dots$. We will express the solution for this semi-infinite chain in terms of the solutions for an infinite chain. Let $T = \sum_{n=-\infty}^{\infty} |n+1\rangle \langle n|$ denote the right-translation operator on the infinite chain. For reasons that will become clear, let us consider a hopping Hamiltonian $\hat{M}$ on the infinite chain, but with the coupling between sites $0$ and $1$ removed:
\begin{equation}
    \hat{M} \coloneqq z \mathds{1} + \beta(T+T^{-1}) - \underbrace{\beta\left(|0\rangle\langle 1| + \mathrm{h.c.}\right)}_{\coloneqq v}.
\end{equation}

\begin{figure}[h]
    \tikzset{every picture/.style={line width=0.75pt}} 
    \begin{tikzpicture}[x=0.75pt,y=0.75pt,yscale=-1,xscale=1]
        
        \draw  [fill={rgb, 255:red, 0; green, 0; blue, 0 }  ,fill opacity=1 ] (127,130) .. controls (127,128.34) and (128.34,127) .. (130,127) .. controls (131.66,127) and (133,128.34) .. (133,130) .. controls (133,131.66) and (131.66,133) .. (130,133) .. controls (128.34,133) and (127,131.66) .. (127,130) -- cycle ;
        \draw  [fill={rgb, 255:red, 0; green, 0; blue, 0 }  ,fill opacity=1 ] (147,130) .. controls (147,128.34) and (148.34,127) .. (150,127) .. controls (151.66,127) and (153,128.34) .. (153,130) .. controls (153,131.66) and (151.66,133) .. (150,133) .. controls (148.34,133) and (147,131.66) .. (147,130) -- cycle ;
        \draw  [fill={rgb, 255:red, 0; green, 0; blue, 0 }  ,fill opacity=1 ] (167,130) .. controls (167,128.34) and (168.34,127) .. (170,127) .. controls (171.66,127) and (173,128.34) .. (173,130) .. controls (173,131.66) and (171.66,133) .. (170,133) .. controls (168.34,133) and (167,131.66) .. (167,130) -- cycle ;
        \draw  [fill={rgb, 255:red, 0; green, 0; blue, 0 }  ,fill opacity=1 ] (187,130) .. controls (187,128.34) and (188.34,127) .. (190,127) .. controls (191.66,127) and (193,128.34) .. (193,130) .. controls (193,131.66) and (191.66,133) .. (190,133) .. controls (188.34,133) and (187,131.66) .. (187,130) -- cycle ;
        \draw  [fill={rgb, 255:red, 0; green, 0; blue, 0 }  ,fill opacity=1 ] (207,130) .. controls (207,128.34) and (208.34,127) .. (210,127) .. controls (211.66,127) and (213,128.34) .. (213,130) .. controls (213,131.66) and (211.66,133) .. (210,133) .. controls (208.34,133) and (207,131.66) .. (207,130) -- cycle ;
        \draw  [fill={rgb, 255:red, 0; green, 0; blue, 0 }  ,fill opacity=1 ] (227,130) .. controls (227,128.34) and (228.34,127) .. (230,127) .. controls (231.66,127) and (233,128.34) .. (233,130) .. controls (233,131.66) and (231.66,133) .. (230,133) .. controls (228.34,133) and (227,131.66) .. (227,130) -- cycle ;
        \draw    (123,130) -- (170,130) ;
        \draw    (190,130) -- (237,130) ;
        
        \draw (187,134) node [anchor=north west][inner sep=0.75pt]  [font=\tiny]  {$1$};
        \draw (207,134) node [anchor=north west][inner sep=0.75pt]  [font=\tiny]  {$2$};
        \draw (227,134) node [anchor=north west][inner sep=0.75pt]  [font=\tiny]  {$3$};
        \draw (167,134) node [anchor=north west][inner sep=0.75pt]  [font=\tiny]  {$0$};
        \draw (142,134) node [anchor=north west][inner sep=0.75pt]  [font=\tiny]  {$-1$};
        \draw (121,134) node [anchor=north west][inner sep=0.75pt]  [font=\tiny]  {$-2$};
        \draw (237,127) node [anchor=north west][inner sep=0.75pt]  [font=\tiny]  {$\cdots $};
        \draw (107,127) node [anchor=north west][inner sep=0.75pt]  [font=\tiny]  {$\cdots $};
    \end{tikzpicture}        
\end{figure}

Since the two semi-infinite halves of the chain are identical and disconnected, we have
\begin{align}
    &(\hat{M}^{-1})_{11} = (\hat{M}^{-1})_{00} = (M^{-1})_{11}, \\
    &(\hat{M}^{-1})_{10} = (\hat{M}^{-1})_{01} = 0.
\end{align}
Define the states $|\pm\rangle = (|0\rangle \pm |1\rangle) / \sqrt{2}$, and let $\hat{G} = \hat{M}^{-1}$ and $G = \left[z \mathds{1} + \beta(T + T^{-1})\right]^{-1}$. Using the previous two equations, we get
\begin{align}
    (M^{-1})_{11} &= \hat{G}_{++} = \langle + | \hat{G} | + \rangle, \nonumber\\
    &= \sum_{n=0}^{\infty} \langle + | (G v)^{n} G |+ \rangle,
\end{align}
where we used $\hat{M} = G^{-1} - v$ and expanded $\left[G^{-1} - v\right]^{-1} = \left[\mathds{1} - G v\right]^{-1} G = \left[\mathds{1} + G v + (G v)^{2} + \cdots\right]G$.

Let $R_{01}$ denote the operator that reflects about the bond $01$. $G$ is reflection symmetric, so $[G, R_{01}] = 0$, while $R_{01} |\pm \rangle = \pm | \pm \rangle$. Together this implies $G_{+-} = G_{-+} = 0$. Then, using $v = \beta(|+\rangle \langle + | - | - \rangle \langle - |)$, one can show by induction that $\langle + | (G v)^{n} G | + \rangle = G_{++}^{n+1} \beta^{n}$. Summing the geometric series, we get
\begin{equation}
    (M^{-1})_{11} = \dfrac{G_{++}}{1 - \beta G_{++}},
\end{equation}
so we have succeeded in expressing the Green's function on the semi-infinite chain in terms of a Green's function on the infinite chain. Expanding the $|+\rangle$ state, $G_{++}$ is given by
\begin{align}
    G_{++} &= \dfrac{1}{2}\left(G_{00} + G_{01} + G_{10} + G_{11}\right)
\end{align}
Since $G$ now involves the simple inversion of a translationally symmetric operator, we can insert a complete set of momentum eigenstates and use $|1\rangle = T |0\rangle$ to obtain
\begin{equation}
    G_{++}(z) = \int_{-\pi}^{\pi} \dfrac{1 + \cos{q}}{z + 2\beta \cos{q}} \dfrac{\diff q}{2\pi}.
\end{equation}
This can be evaluated using residue calculus; defining
\begin{equation}
    u_{\pm}(z) \coloneqq \dfrac{-z \pm \sqrt{z^{2} - 4\beta^{2}}}{2\beta},
\end{equation}
we get
\begin{equation}
    \dfrac{G_{++}(z)}{1-\beta G_{++}(z)} = \begin{dcases}
        \dfrac{-u_{+}(z)}{\beta}, & |u_{+}(z)| < 1,\\
        \dfrac{-u_{-}(z)}{\beta}, & |u_{-}(z)| < 1.
    \end{dcases}
\end{equation}
Note that $u_{+} u_{-} = 1$, so we are either in one of these cases, or the integral is undefined. We obtain the final result for the Green's function by restoring $z \mapsto z + i\Gamma$ in $G_{++}$, so that
\begin{align}
    \oprod{\wt{\sigma}^{z}_{k}(z)}{\wt{\sigma}^{z}_{k}}_{\mu} &= \frac{1}{2} \oprod{a_{k,0,+}(z)}{a_{k,0,+}}_{\mu}, \nonumber \\
    &= \dfrac{i}{z + 2 v_{k}^{2} \frac{G_{++}(z + i \Gamma)}{1 - \beta G_{++}(z + i \Gamma)}}, \nonumber \\
    &= \begin{dcases}
        \dfrac{i}{z + \eta^{-2}\left(-z - i\Gamma + \sqrt{(z+i\Gamma)^{2} - 4 \eta^{2} \sin^{2}(k/2)}\right)}, & |u_{+}(z+i\Gamma)| < 1,\\
        \dfrac{i}{z + \eta^{-2}\left(-z - i\Gamma - \sqrt{(z+i\Gamma)^{2} - 4 \eta^{2} \sin^{2}(k/2)}\right)}, & |u_{-}(z+i\Gamma)| < 1,
    \end{dcases}
\end{align}
where we used $v_{k} = i \sin(k/2)$ and $\beta = i \eta v_{k}$. One can check from this solution that as $k \to 0$ and $z \to 0$, the Green's function has a diffusive pole, $\oprod{\wt{\sigma}^{z}_{k}(z)}{\wt{\sigma}^{z}_{k}}_{\mu} \approx i/(z + i D k^{2})$, with $D = (2\Gamma)^{-1}$ as claimed.

\subsection{Real time correlation function}
In this section we obtain the real time $k$-space correlation function by performing the inverse Laplace transform of the result for $\oprod{\wt{\sigma}^{z}_{k}(z)}{\wt{\sigma}^{z}_{k}}_{\mu} \equiv C(z)$ obtained in the previous section. We start by re-expressing the Green's function as a contour integral. Noting that it can be written as $i / (z + 2\beta \eta^{-2} u_{\mathrm{min}})$, where $u_{\mathrm{min}}$ is the smaller in magnitude of $u_{\pm}(z+i \Gamma)$, we can write
\begin{align}
    C(z) &= \oint_{|u|=1} \dfrac{\diff u}{2\pi i} \dfrac{i}{z + 2 \beta \eta^{-2} u}\left[\dfrac{1}{u - u_{+}(z+i \Gamma)} + \dfrac{1}{u - u_{-}(z+i \Gamma)} - \mathrm{Res}\left(\dfrac{1}{u - u_{+}(z+i \Gamma)} + \dfrac{1}{u - u_{-}(z+i \Gamma)} ; u = \dfrac{-z}{2\beta \eta^{-2}}\right)\right],\\
    &= \oint_{|u|=1} \dfrac{\diff u}{2\pi i} \dfrac{i}{z + 2 \beta \eta^{-2} u}\left[ \dfrac{z + i \Gamma + 2\beta u}{u \left[z + i \Gamma + \beta(u + u^{-1})\right]} - \dfrac{4\beta(i \Gamma + (1-\eta^{2})z)}{\eta^{2}(\eta^{2}-2)z^{2} - 2i \Gamma \eta^{2} z + 4 \beta^{2}}\right].
\end{align}
In order to avoid the spurious contribution from the pole of $1/(z+2\beta \eta^{-2} u)$, we have subtracted off a constant in the square brackets to set the residue to zero. The advantage of this contour integral formulation is that the denominators are now all simple rational functions of $z$, so we can perform the inverse Laplace transform, $C(t) = (1/2\pi) \int_{-\infty + i \gamma}^{\infty + i\gamma} e^{-i z t} C(z) \diff z$, using residue calculus. The value of $\gamma$ is taken large enough that all poles of $C(z)$ lie below the integration contour, and for $t > 0$ we evaluate the integral by closing the contour in the lower-half plane. There are three relevant poles: one at $z = -i \Gamma - \beta(u + u^{-1})$, and a pair at
\begin{equation}
    z_{\pm} \coloneqq \dfrac{i}{2-\eta^{2}}\left(-\Gamma \pm \sqrt{\Gamma^{2} - 4 \beta^{2}(2 \eta^{-2}-1)}\right).
\end{equation}
We will split the correlation function as $C(t) \equiv C_{1}(t) + C_{2}(t)$, corresponding to the contribution from the pole at $z = -i \Gamma - \beta(u + u^{-1})$, and the joint contrbution from $z_{\pm}$ respectively. Note that the same subtraction that previously set the residue to zero of the pole in the $u$-integral at $z + 2 \beta \eta^{-2} u = 0$ now also sets the residue to zero of the pole in the $z$-integral with the same condition $z + 2 \beta \eta^{-2} u = 0$, so we do not need to worry about its contribution.

The pole at $z = -i \Gamma - \beta(u + u^{-1})$ gives rise to the ballistic term in the correlation function, given by
\begin{align}
    C_{1}(k,t) = e^{-\Gamma t} \int_{-\pi}^{\pi} \dfrac{\diff q}{2\pi} \dfrac{2 \beta \eta^{2} \sin(q) e^{2 i \beta t \cos{q}}}{-\Gamma \eta^{2} + 2 \beta \sin{q} - (1-\eta^{2})2i \beta \cos{q}}.
\end{align}
To see that this corresponds to ballistic physics, note that, in the $|\mu| \to \infty$ limit, we have $\Gamma \to 0$ and $\eta^{2} \to 1$, in which case this reduces to the integral expression for the single-particle solution $C_{1}(k,t) = J_{0}(2t \sin(k/2))$, where $J_{0}(z) = \int_{-\pi}^{\pi} \frac{\diff q}{2\pi} e^{i z \cos{q}}$ is a Bessel function.

The other contribution $C_{2}(k,t)$ from the poles $z_{\pm}$ gives rise to diffusion; its general expression is
\begin{equation}
    C_{2}(k,t) = \dfrac{2}{2 - \eta^{2}} \dfrac{1}{z_{+} - z_{-}}\left[e^{-i z_{+} t}(i \Gamma + (1-\eta^{2})z_{+}) \, \mathcal{I}\left(\left|\dfrac{z_{+}}{2 \beta \eta^{-2}}\right|<1\right) - e^{-i z_{-} t}(i \Gamma + (1-\eta^{2})z_{-}) \, \mathcal{I}\left(\left|\dfrac{z_{-}}{2 \beta \eta^{-2}}\right|<1\right)\right],
\end{equation}
where $\mathcal{I}$ is an indicator function. The values of the indicator functions are set by the momentum through the parameter $\beta = \tanh(\mu/2) \sin(k/2)$. We define the auxiliary function
\begin{equation}
    g_{k} \coloneqq \sqrt{1 - (2-\eta^{2}) (2/\Gamma)^{2} \sin^{2}(k/2)}.
\end{equation}
Then for small momenta, meaning $\sin^{2}(k/2) < (\eta \Gamma / 2)^{2}$, we have
\begin{align}
    C_{2}(k,t) &= \dfrac{1}{2-\eta^{2}}\left(c_{\mu}^{-2} + g_{k}^{-1}\right) \exp\left[-\dfrac{\Gamma t}{2-\eta^{2}} \left(1 - g_{k}\right)\right].
\end{align}
As $k \to 0$, one obtains the diffusive form $\exp(-D k^{2} t)$ with $D = (2\Gamma)^{-1}$. Note that this small momentum regime is exponentially small in $\mu$, given the scaling $D \sim \exp[\mathcal{O}(\mu)]$. For larger momenta, $\sin^{2}(k/2) > (\eta \Gamma / 2)^{2}$, one instead has 
\begin{align}
    C_{2}(k,t) = \dfrac{2 e^{-\frac{\Gamma t}{2-\eta^{2}}}}{2-\eta^{2}}\left[ c_{\mu}^{-2} \cosh\left(\frac{g_{k}\Gamma t}{2-\eta^{2}}\right) + g_{k}^{-1}\sinh\left(\frac{g_{k}\Gamma t}{2-\eta^{2}}\right)\right].
\end{align}
This discontinuity in $C_{2}$ at $\sin^{2}(k/2) = (\eta \Gamma / 2)^{2}$ is compensated by a discontinuity in $C_{1}$, such that $C(k,t)$ is a smooth function of $k$. There is a further change at even larger momenta, $\sin^{2}(k/2) > \dfrac{1}{2-\eta^{2}} (\Gamma/2)^{2}$, where $g_{k}$ becomes imaginary; in this region the expression for $C_{2}$ is obtained using $\cosh(i x) = \cos(x)$ and $\sinh(i x) = i \sin(x)$.

\section{Operator weight distribution}
In this section we build intuition about the action of the DAOE$_\mu$ dissipator by considering its effect on the operator weight distribution. This is relevant for understanding the effect of the dissipation on the operator entanglement entropy; since there are only $\binom{L}{l} 3^{l}$ independent spin-$\frac{1}{2}$ operators of weight $l$, if the operator distribution is biased towards small $l$ by DAOE$_\mu$, it is reasonable to expect that one should be able to describe it faithfully using less memory than without the application of DAOE$_\mu$. We will consider the operator weight in both the original Pauli basis and the new $\wt{\sigma}$ basis defined in \cref{eq:tilde_pauli_def} of the main text. The former is relevant for the operator entanglement entropy, since traditionally the Schmidt decomposition is performed with respect to the Hilbert-Schmidt product, for which the original Paulis form an orthonormal operator basis.

Given an operator $O$, its component on weight-$l$ $\wt{\sigma}$ strings is defined as
\begin{equation}
    w_{l}(\mu) \coloneqq \sum_{\bm{\alpha} : |\bm{\alpha}| = l} \left|\oprod{\wt{\sigma}^{\bm{\alpha}}}{O}_{\mu}\right|^{2},
\end{equation}
where $\wt{\sigma}^{\bm{\alpha}} = \otimes_{i=1}^{L} \wt{\sigma}^{\alpha_{i}}$, $\alpha_{i} \in \{0,x,y,z\}$, and $|\bm{\alpha}| = |\{i : \alpha_{i} \neq 0\}|$. The original Pauli weight corresponds to $\mu = 0$. Since the $\wt{\sigma}$ strings form an orthonormal basis with respect to the thermal inner product, we have $\sum_{l=0}^{L} w_{l}(\mu) = \oprod{O}{O}_{\mu}$.

\subsection{Random phase Pauli strings}
As the operator $O(t)$ evolves in time, it will develop an overlap with Pauli strings of higher weight. For an analytically tractable example, we consider an operator which is a product of random single-site Pauli operators,
\begin{equation}
    O \propto \otimes_{i=1}^{L} \left(\mathds{1} + e^{i \theta_{z,i}} Z_{i} + e^{i \theta_{x,i}} X_{i} + e^{i \theta_{y,i}} Y_{i}\right),
\end{equation}
where $Z_{i} \equiv \sigma^{z}_{i}$, etc., and the phases $\theta_{\alpha,i}$ are i.i.d.\ random variables drawn from the uniform distribution on $[0,2\pi]$. We chose a fixed operator basis because the Heisenberg picture operator dynamics is independent of $\mu$, with only the operator inner product changing with $\mu$. One potentially important ingredient missing from this simple example is the effect of conservation laws: if $\sigma^{z}_{\mathrm{tot}}$ is conserved then at long times we expect the operator dynamics to be dominated by the conserved modes, so that $\sigma^{z}$ would likely have a larger weight in the decomposition of a time-evolved operator than $\sigma^{x,y}$. Nonetheless, we can learn some basic lessons from this simple model. In terms of the new operator basis, we have
\begin{equation}
    O = \bigotimes_{i=1}^{L} \left(\dfrac{(1 - t e^{i \theta_{z,i}}) \mathds{1} + c^{-1} e^{i \theta_{z,i}} \wt{Z} + c^{-1/2} e^{i \theta_{i,x}} \wt{X} + c^{-1/2} e^{i \theta_{i,y}} \wt{Y}}{\sqrt{|1 - t e^{i \theta_{z,i}}|^{2} + c^{-2} + 2 c^{-1}}}\right),
\end{equation}
where again $\wt{Z}_{i} \equiv \wt{\sigma}^{z}_{i}$, etc., $t \equiv \tanh(\mu/2)$, $c \equiv \cosh(\mu/2)$, and we have now included the normalization factor (w.r.t.\ the thermal inner product). Averaging over the random phases, the averaged overlap with weight $l$ operators is then
\begin{equation}
    \overline{w_{l}(\mu)} = \binom{L}{l} \overline{\left(\dfrac{|1 - t e^{i \theta_{z}}|^{2}}{|1 - t e^{i \theta_{z}}|^{2} + c^{-2} + 2 c^{-1}}\right)}^{L-l} \sum_{k=0}^{l} \binom{l}{k} c^{-2k} \left(\dfrac{2}{c}\right)^{l-k} \overline{\left(\dfrac{1}{|1 - t e^{i \theta_{z}}|^{2} + c^{-2} + 2 c^{-1}}\right)}^{l}.
\end{equation}
The prefactor to the sum comes from choosing the $l$ sites for the non-identity operators, while the sum over $k$ comes from choosing how many of these non-identities will be $\wt{Z}$ operators. For the uniform distribution, the averages can be computed using a residue calculation, giving
\begin{align}
    \overline{\left(\dfrac{1}{|1 - t e^{i \theta_{z}}|^{2} + c^{-2} + 2 c^{-1}}\right)} &= \dfrac{1}{4} \dfrac{\cosh(\mu/2)}{\cosh(\mu/4)},\\
    \overline{\left(\dfrac{|1 - t e^{i \theta_{z}}|^{2}}{|1 - t e^{i \theta_{z}}|^{2} + c^{-2} + 2 c^{-1}}\right)} &= 1 - \sinh(\mu/4) \left[\cosech(\mu) + \cosech(\mu/2)\right],
\end{align}
and so, after performing the sum over $k$, we get a binomial distribution
\begin{equation}
    \overline{w_{l}(\mu)} = \binom{L}{l} \left[ 1 - p(\mu)\right]^{L-l} p(\mu)^{l},
    \label{eq:wl_mu_prediction}
\end{equation}
where the probability $p(\mu)$ to have a non-identity on a given site is
\begin{equation}
p(\mu) \coloneqq \sinh(\mu/4) \left[\cosech(\mu) + \cosech(\mu/2)\right].
\end{equation}
As $\mu \to 0$ we have $p(\mu) \to 3/4$ as expected. The average operator weight $\overline{\langle s(\mu) \rangle} = \sum_{l=0}^{L} l \overline{w_{l}(\mu)}$ is then given by
\begin{equation}
    \overline{\langle s(\mu) \rangle} = p(\mu) L.
\end{equation}
The linear scaling with $L$ is a consequence of the fact that we are averaging over products of random single-site Paulis.

\subsection{Effect of DAOE$_\mu$ on operator weight distribution}
\label{sec:operator_weight}
We now consider the effect of performing one round of DAOE$_\mu$ on the operator weight distribution of an operator which is initially a random phase Pauli string. For simplicity we consider $\gamma = \infty$, and fix $\ell_{*} = 3$. We average the results over 1000 random phase Pauli strings. For each sample, we perform DAOE$_\mu$, put the dissipated operator back into canonical form (so it is normalized w.r.t.\ the Hilbert-Schmidt product), and calculate the components on different weights. With the operator represented as an MPS in a doubled Hilbert space, the component on weight $l$ operators can be extracted using a `weight $l$ superoperator' represented by an MPO of bond dimension $l+2$. We will first describe this for the standard Pauli basis, with the generalization to the $\wt{\sigma}$ Pauli basis being straightforward. As with the MPO for the DAOE superoperator, we label the local basis states by $n=\mathds{1},X,Y,Z$, and then write the local MPO tensor $W_{ab}^{n n^{\prime}}$ as a matrix acting on the virtual indices $a,b=0,1,\dots,l+1$. The nonzero entries are $W^{\mathds{1} \mathds{1}}_{ab} = \delta_{ab}$ and $W^{XX}_{ab} = W^{YY}_{ab} = W^{ZZ}_{ab} = \delta_{a=b-1}$. We contract this MPO with the vectors $v_{L} = (1,0,0,\dots)$ on the left boundary and $v_{R} = (0,0,\dots,0,1,0)$ on the right, where the $1$ in $v_{R}$ is at index $b=l$, so that only the weight $l$ component of an operator gives a nonzero contribution. The component of operator $O$ on weight $l$ Paulis is then given by the expectation value of this MPO superoperator with respect to the MPS representing $O$. To instead calculate the weight with respect to the $\wt{\sigma}$ basis, we simply perform an onsite basis transformation between the standard Pauli basis and the $\wt{\sigma}$ basis before and after applying the weight superoperator, as we did when applying the DAOE$_\mu$ dissipator.

\begin{figure}[t]
    \centering
    \subfigimg[width=0.4\linewidth]{(a)}{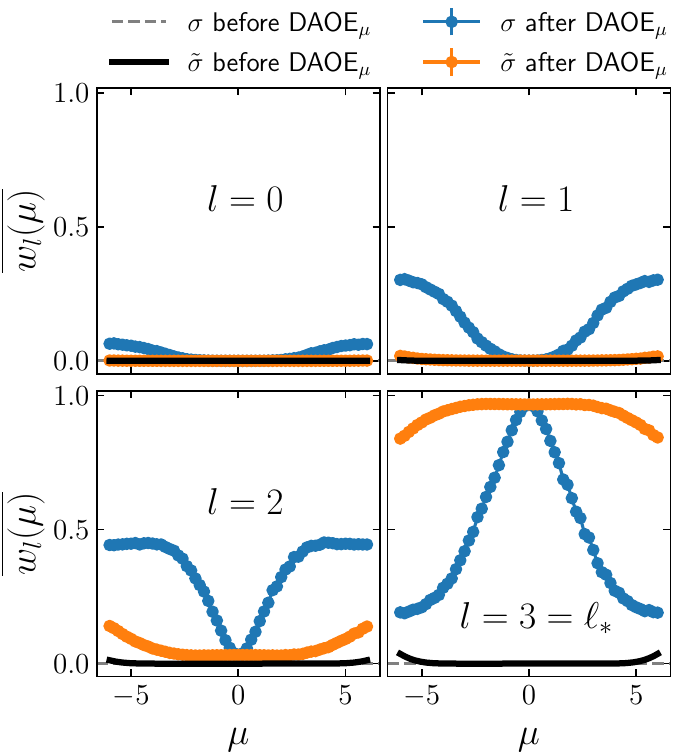}
    \hspace{4em}
    \subfigimg[width=0.4\linewidth]{(b)}{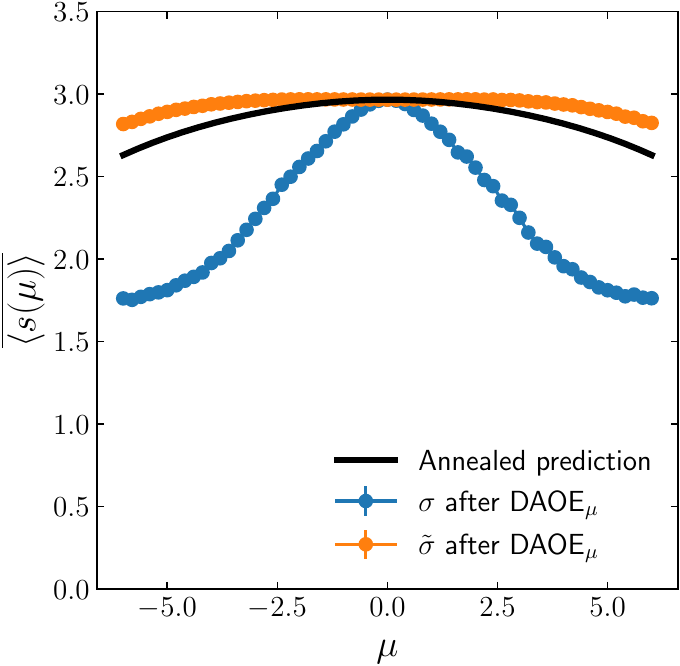}
    \caption{Effect of DAOE$_\mu$ on the operator weight distribution of random phase Pauli strings. We apply DAOE$_\mu$ with $\gamma = \infty$ and $\ell_{*} = 3$, and average over 1000 operator samples in a system of size $L=32$. \textbf{(a)} Averaged weight $\overline{w_{l}(\mu)}$ on operators of different weights $l$, before and after applying DAOE$_\mu$. The `before DAOE$_\mu$' results are given by $\binom{L}{l} 3^{l} 4^{-L}$ for the $\sigma$ basis and by \cref{eq:wl_mu_prediction} for the $\wt{\sigma}$ basis. The `after DAOE$_\mu$' results are calculated using the weight superoperator MPO described in the text. \textbf{(b)} The average operator weight $\overline{\langle s(\mu)\rangle}$ in the $\sigma$ and $\wt{\sigma}$ bases. We see that the average weight in the $\sigma$ basis is less than or equal to that in the $\wt{\sigma}$ basis, such that, for the purposes of operator entanglement, DAOE$_\mu$ can be thought of as performing $(\mu=0)$DAOE with a $\mu$-dependent effect cutoff $\ell_{*, \mathrm{eff}}(\mu)$ less than or equal to the true $\ell_{*}$.}
    \label{fig:operator_weight_distribution}
\end{figure}

We show the results of this procedure in \cref{fig:operator_weight_distribution} for a system of size $L=32$. In \cref{fig:operator_weight_distribution}(a) we show the averaged weight $\overline{w_{l}(\mu)}$ on operators of different weights $l$, before and after applying DAOE$_\mu$. Since we are using $\gamma = \infty$, after performing DAOE$_\mu$ the maximum operator length is strictly $l = \ell_{*} = 3$. The `before DAOE$_\mu$' results are given by $\binom{L}{l} 3^{l} 4^{-L}$ for the $\sigma$ basis and by \cref{eq:wl_mu_prediction} for the $\wt{\sigma}$ basis, while the `after DAOE$_\mu$' results are calculated using the weight superoperator MPO. Since the system size $L =32$ is much larger than the cutoff $\ell_{*} = 3$, before DAOE$_\mu$ the component on small operators with $l \leq \ell_{*}$ is fairly negligible, but after DAOE$_\mu$ all the operator weight is redistributed on to these small operators. In the $\wt{\sigma}$ basis, most of the weight after DAOE$_\mu$ is concentrated near $l = \ell_{*} = 3$ simply because there are many more possible $l = 3$ operators than $l < 3$ operators.

Transforming back to the original $\sigma$ basis, the weight from these $l = 3$ operators is partly redistributed to $l < 3$ operators (with respect to the $\sigma$ basis), effectively because the $\wt{\sigma}^{z}$ operator contains an identity term. In more detail, apart from overall normalizations by factors of $\cosh(\mu/2)$, the main difference between the original Pauli basis and the $\wt{\sigma}$ Pauli basis is in the new $\sigma^{z}$ operator, $\wt{\sigma}^{z} = c_{\mu}(\sigma^{z} - \langle \sigma^{z} \rangle_{\mu} \mathds{1})$, which now contains an identity term. One consequence of this is that the average operator weight, $\langle s(\mu) \rangle = \sum_{l=0}^{L} l w_{l}(\mu)$, is always lower in the original Pauli basis than in the new basis, i.e.\ $\langle s(\mu = 0)\rangle \leq \langle s(\mu)\rangle$. For example, if we consider an operator string $\wt{\sigma}^{z}_{i} \wt{\sigma}^{z}_{j} \wt{\sigma}^{z}_{k}$ which has weight 3 in the $\wt{\sigma}$ basis, expanding out the operator in the Pauli basis, $\wt{\sigma}^{z}_{i} \wt{\sigma}^{z}_{j} \wt{\sigma}^{z}_{k} = c_{\mu}^{3} (\sigma^{z} - \langle \sigma^{z} \rangle_{\mu} \mathds{1})_{i} (\sigma^{z} - \langle \sigma^{z} \rangle_{\mu} \mathds{1})_{j} (\sigma^{z} - \langle \sigma^{z} \rangle_{\mu} \mathds{1})_{k}$, we see that it becomes a superposition of operators with $\mu=0$ weight of less than or equal to 3, so its average $\mu = 0$ weight is less than 3 provided the coefficient $\langle \sigma^{z}\rangle_{\mu}$ is nonzero. Accordingly, in \cref{fig:operator_weight_distribution}(b) we show the average operator weight $\overline{\langle s(\mu) \rangle}$ in the $\sigma$ and $\wt{\sigma}$ bases. As expected, we see that the average weight in the $\sigma$ basis is less than or equal to that in the $\wt{\sigma}$ basis, such that, for the purposes of operator entanglement, DAOE$_\mu$ can be thought of as performing $(\mu=0)$DAOE with a $\mu$-dependent effective cutoff $\ell_{*, \mathrm{eff}}(\mu)$ less than or equal to the true $\ell_{*}$. From \cref{fig:operator_weight_distribution}(b), we see that $\ell_{*, \mathrm{eff}}(\mu)$ becomes increasingly smaller than the true $\ell_{*}$ as $|\mu|$ increases.

In terms of the $\wt{\sigma}$ basis, the average weight after DAOE$_\mu$ is close to $l = \ell_{*} = 3$, again simply because there are many more $l = 3$ operators than $l < 3$ operators. We also show the prediction from the `annealed average' $\sum_{l=0}^{\ell_{*}} l \overline{w_{l}(\mu)} / \sum_{l=0}^{\ell_{*}} \overline{w_{l}(\mu)}$ where we separately average the mean weight and the mean normalization, and $\overline{w_{l}(\mu)}$ is given by \cref{eq:wl_mu_prediction}; in the numerics we instead normalize the operator before averaging over samples. The annealed average comes reasonably close to the quenched average from the numerics, with the agreement better at smaller $\mu$.

\section{Failure of the DAOE${}_{0}$ dissipator at low filling}
\begin{figure}[h]
    \centering
    \subfloat[$\mu = 0.5$.]{\includegraphics[width=0.33\linewidth]{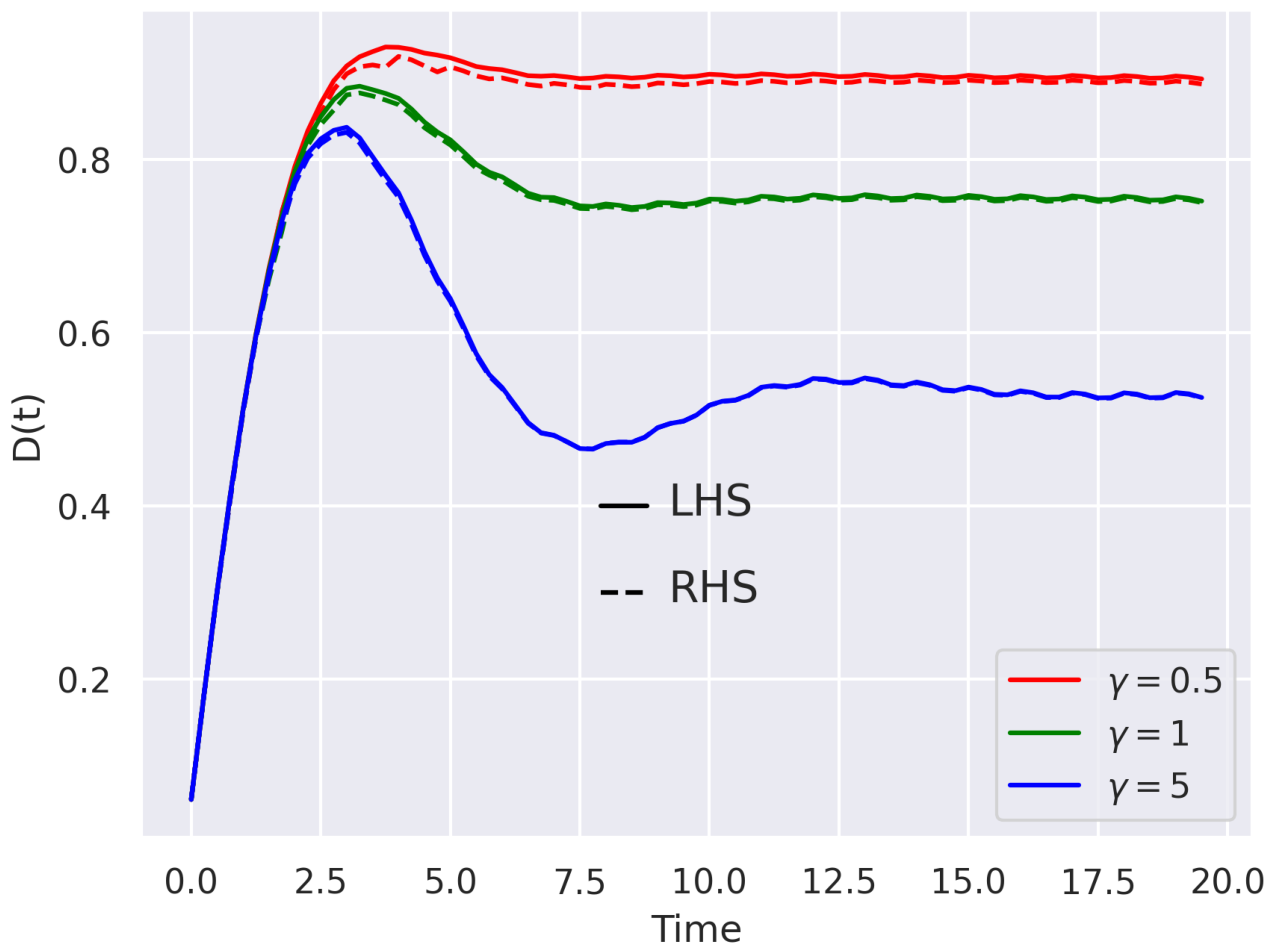}}
    \hfill
    \subfloat[$\mu = 2$.]{\includegraphics[width=0.33\linewidth]{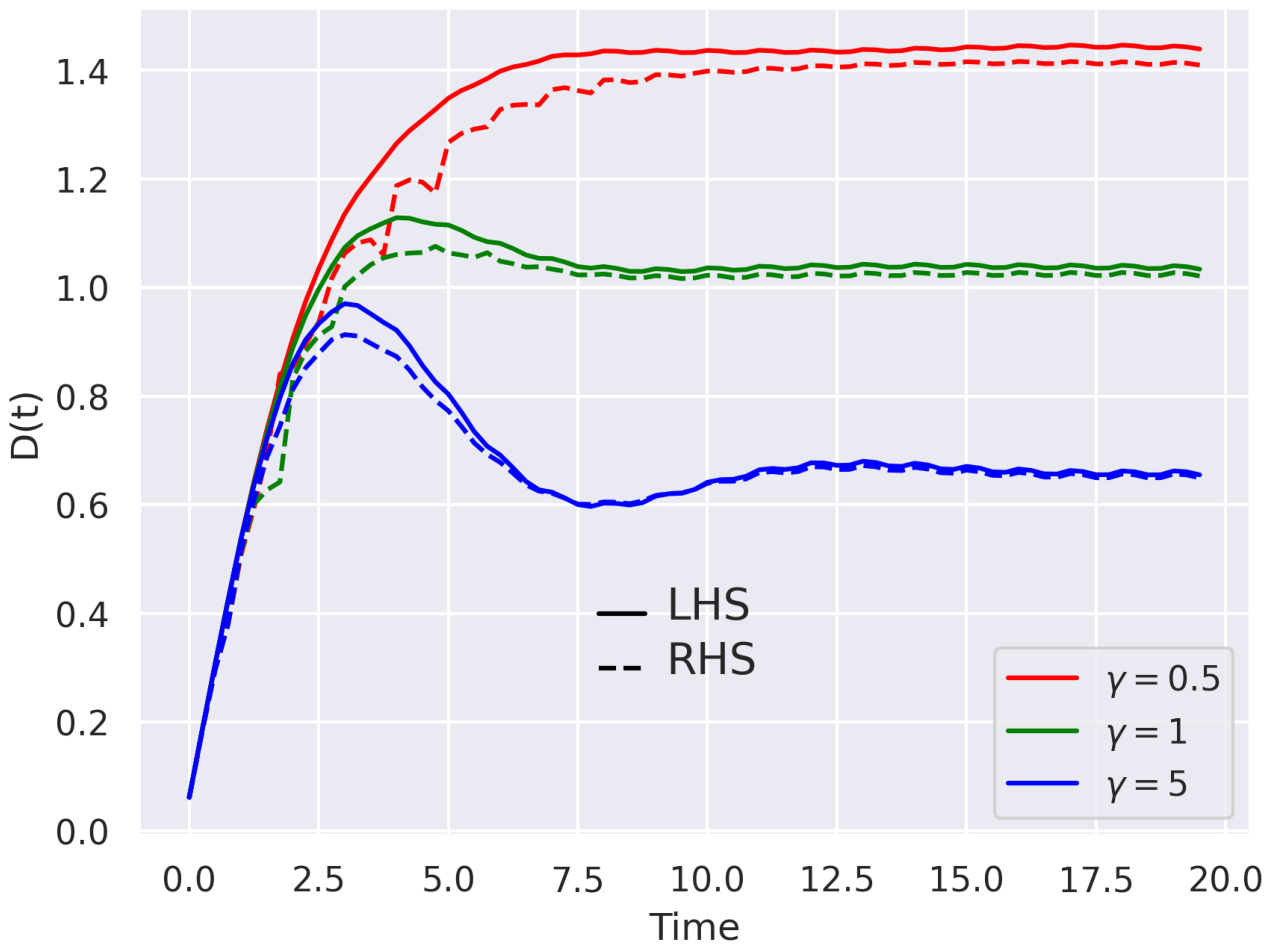}}
    \hfill
    \subfloat[$\mu = 6$.]{\includegraphics[width=0.33\linewidth]{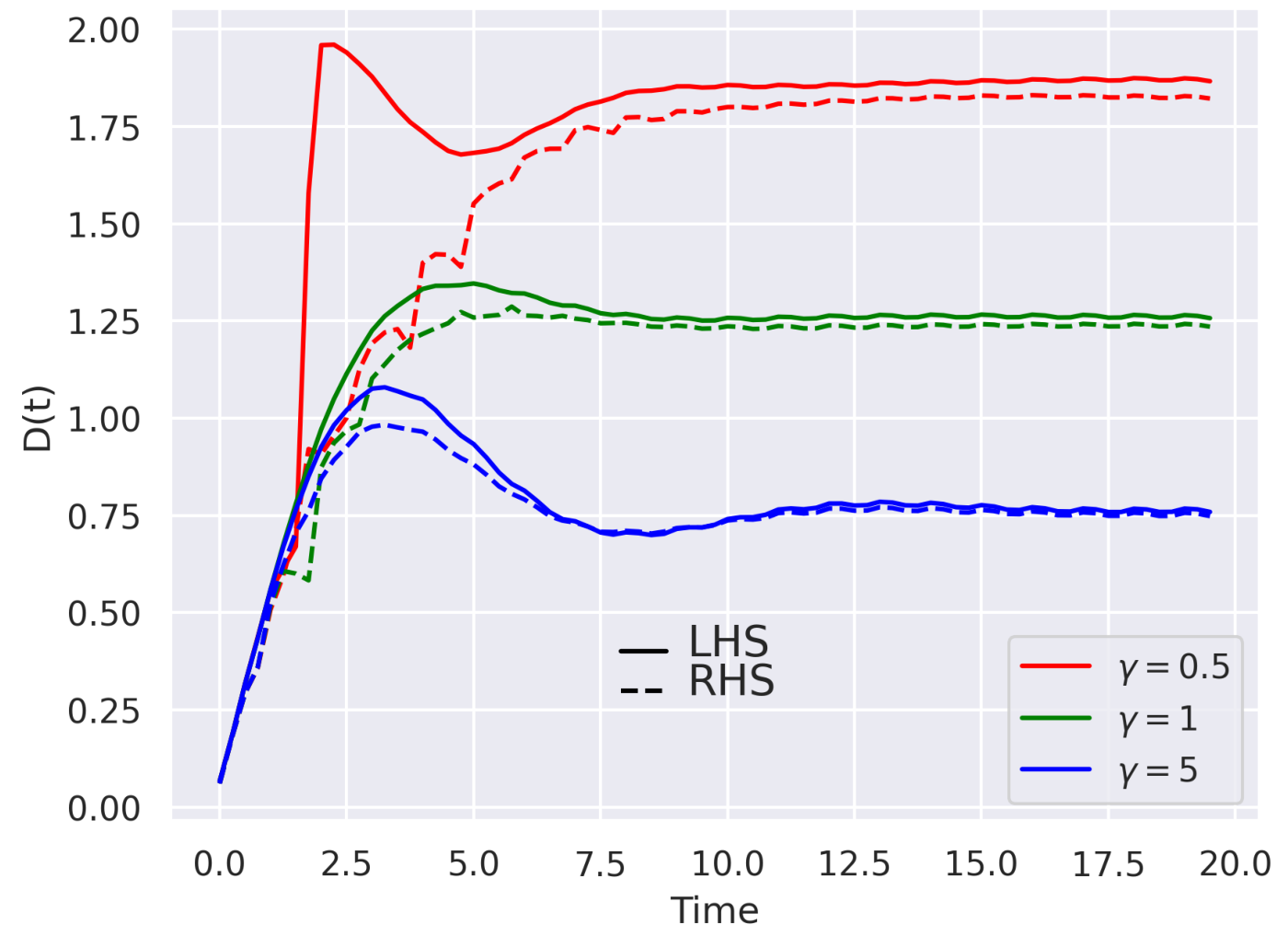}}
    \caption{A comparison between the LHS and RHS of \cref{eq:correlation_func_eta_expansion}, demonstrating how the DAOE$_0$ dissipator can result in diffusion even when we expect nondiffusive transport. These figures show how the agreement between the numerically exact DAOE evolution and the $\eta$ expansion becomes better at small $\mu$. All figures use $\ell_{*} = 3$ and $\chi_{\mathrm{max}} = 128$.}
    \label{fig:daoe_default_diffusion}
\end{figure}

In this section we substantiate the argument given in the main text for why the original DAOE${}_{0}$ dissipator does not give the correct scaling for the diffusion constant as $|\mu| \to \infty$. In the text surrounding \cref{eq:correlation_func_eta_expansion}, we argued that, at least for $\gamma = \infty$, only keeping Pauli strings up to any \textit{finite} cutoff $\ell_{*}$ would result in a finite diffusion constant even as $|\mu| \to \infty$, whereas we expect that $D(\mu) \sim e^{|\mu|}$ should diverge exponentially. To check this argument away from the $\gamma = \infty$ limit, we show in \cref{fig:daoe_default_diffusion} a comparison between the correlation function computed using the full DAOE$_0$ evolution at finite $\gamma$ (LHS), and the approximation from \cref{eq:correlation_func_eta_expansion} (RHS). While there are some significant deviations at early times, for later times the agreement is remarkably close. The agreement generally improves for smaller $\mu$ and larger $\gamma$, as expected. 

\section{Convergence of the diffusion constant}
\begin{figure}[h]
    \centering
    \includegraphics[width=0.5\columnwidth]{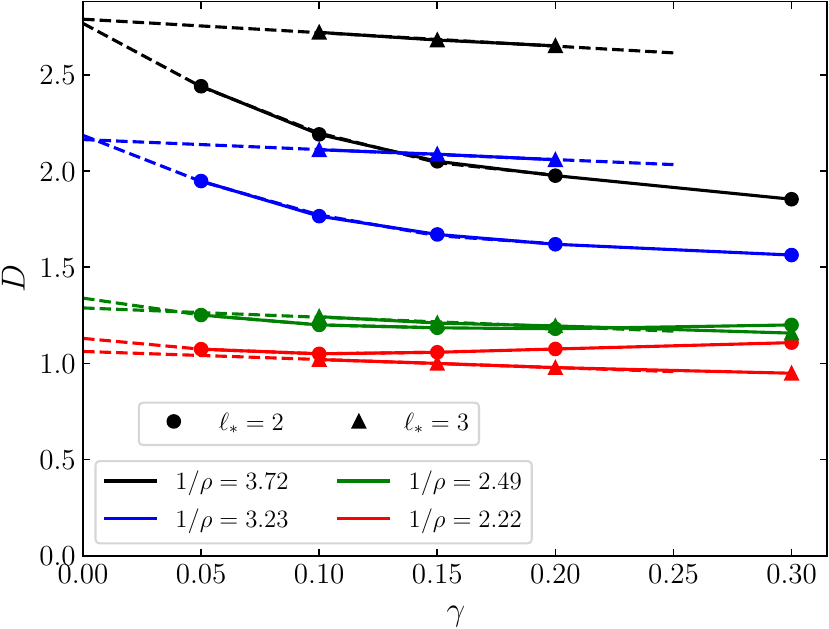}
    \caption{Convergence of the estimated diffusion constants as the dissipation strength $\gamma \to 0$, for different cutoff lengths $\ell^{*}$. The dashed lines show quadratic fits for $\ell^{*} = 2$ and linear fits for $\ell^{*} = 3$, which agree as $\gamma \to 0$ to within $\sim 10\%$.}
    \label{fig:dvsgam}
\end{figure}

By introducing artificial dissipation, we are able to perform controlled simulations of the operator dynamics for long times using modest bond dimension, allowing us to extract a diffusion constant $D(\ell_{*},\gamma)$ which depends on the DAOE cutoff length $\ell_{*}$ and dissipation strength $\gamma$. Ultimately we are interested in the diffusion constant of the original model, which can be recovered in either of two limits: $\ell_{*} \to \infty$, or $\gamma \to 0$. In \cref{fig:dvsgam} we show an extrapolation of the diffusion constant of the XX ladder as $\gamma \to 0$ for a range of different densities $\rho$. We find that as $\gamma \to 0$ the extrapolations for $\ell_{*} = 2$ and $\ell_{*} = 3$ are consistent to within approximately $10 \%$ in the worst case, indicating that we can estimate the true diffusion constant to a relatively good precision. To get precision $\epsilon$ from standard time evolution techniques like TEBD \textit{without} artificial dissipation would require memory resources of roughly $\chi \sim \exp[\mathcal{O}(1/\epsilon^{2})]$, assuming a simulation time of $\mathcal{O}(1/\epsilon^{2})$ and the operator entanglement entropy growing linearly in time~\cite{vonkeyserlingkOperatorBackflowClassical2022}. In practice this is prohibitively large for the timescales required to get converged estimates of diffusion constants in this model at low fillings, so our usage of artificial dissipation was crucial to reach convergence. 

\end{document}